\documentclass[sigconf]{acmart}
\AtBeginDocument{%
  }

\setcopyright{acmlicensed}
\copyrightyear{2018}
\acmYear{2018}
\acmDOI{XXXXXXX.XXXXXXX}
\acmConference[Conference acronym 'XX]{Make sure to enter the correct
  conference title from your rights confirmation email}{June 03--05,
  2018}{Woodstock, NY}
\acmISBN{978-1-4503-XXXX-X/2018/06}

\usepackage{tabularx}
\usepackage{graphicx}
\usepackage{subcaption}
\usepackage{multirow}
\usepackage{enumitem}
\usepackage{appendix}
\usepackage{booktabs}
\usepackage{url}  
\usepackage{hyperref}
\usepackage{colortbl}

\newcommand{\revise}[1]{\textcolor{black}{#1}}

\copyrightyear{2026}
\acmYear{2026}
\setcopyright{cc}
\setcctype{by-nc-nd}
\acmConference[CHI EA '26]{Extended Abstracts of the 2026 CHI Conference on Human Factors in Computing Systems}{April 13--17, 2026}{Barcelona, Spain}
\acmBooktitle{Extended Abstracts of the 2026 CHI Conference on Human Factors in Computing Systems (CHI EA '26), April 13--17, 2026, Barcelona, Spain}
\acmDOI{10.1145/3772363.3798613}
\acmISBN{979-8-4007-2281-3/2026/04}

\begin{document}




\title[Seeing the Reasoning]{Seeing the Reasoning: How LLM Rationales Influence User Trust and Decision-Making in Factual Verification Tasks}









\author{Xin Sun}
\affiliation{%
  \department{Digital Content and Media Sciences Research Division}
  \institution{National Institute of Informatics (NII)}
  \city{Tokyo}
  \country{Japan}
}
\affiliation{%
  \department{Social and Behavioural Science}
  \institution{University of Amsterdam}
  \city{Amsterdam}
  \country{Netherlands}
}
\email{xsun@nii.ac.jp}

\author{Shu Wei}
\affiliation{%
  \department{XR Pediatrics}
  \institution{Yale School of Medicine}
  \city{New Haven}
  \state{Connecticut}
  \country{USA}
}
\email{weishu140@gmail.com}

\author{Jos A Bosch}
\affiliation{%
  \department{Psychology}
  \institution{University of Amsterdam}
  \city{Amsterdam}
  \country{Netherlands}
}
\email{j.a.bosch@uva.nl}

\author{Isao Echizen}
\affiliation{%
  \department{Information and Society Research Division}
  \institution{National Institute of Informatics}
  \city{Tokyo}
  \country{Japan}
}
\email{iechizen@nii.ac.jp}

\author{Saku Sugawara}
\affiliation{%
  \department{Digital Content and Media Sciences Research Division}
  \institution{National Institute of Informatics (NII)}
  \city{Tokyo}
  \country{Japan}
}
\email{saku@nii.ac.jp}

\author{Abdallah El Ali}
\affiliation{%
  \institution{Centrum Wiskunde \& Informatica (CWI)}
  \city{Amsterdam}
  \country{Netherlands}
}
\affiliation{%
  \institution{Utrecht University}
  \city{Utrecht}
  \country{Netherlands}
}
\email{aea@cwi.nl}



\begin{abstract}
Large Language Models (LLMs) increasingly show reasoning rationales alongside their answers, turning ``reasoning’' into a user-interface element. While step-by-step rationales are typically associated with model performance, how they influence users’ trust and decision-making in factual verification tasks remains unclear. We ran an online study (N=68) manipulating three properties of LLM reasoning rationales: presentation format (instant vs. delayed vs. on-demand), correctness (correct vs. incorrect), and certainty framing (none vs. certain vs. uncertain). We found that correct rationales and certainty cues increased trust, decision confidence, and AI advice adoption, whereas uncertainty cues reduced them. Presentation format did not have a significant effect, suggesting users were less sensitive to how reasoning was revealed than to its reliability. Participants indicated they use rationales to primarily audit outputs and calibrate trust, where they expected rationales in stepwise, adaptive forms with certainty indicators. Our work shows that user-facing rationales, if poorly designed, can both support decision-making yet miscalibrate trust.
\end{abstract}



\maketitle


\section{Introduction}

Large language models (LLMs), especially large reasoning models (LRMs)~\cite{lrm_survey}, increasingly show step-by-step reasoning rationales alongside their answers. 
Model-centric work studies these reasoning techniques (e.g., chain-of-thought) mainly for improving model performance~\citep{cot}. 
However, such rationales become \emph{user-facing} as part of the user interface since they are read and interpreted by users, not just internally by models~\citep{users_perceive_cot}.
Yet it remains unclear how showing rationales influence users’ trust and decision-making. \revise{This gap matters because LLM rationales can be fluent and persuasive but incorrect or inconsistent, which prior work defines as unfaithful~\citep{reasoning_incorrect_1,reasoning_incorrect_2,reasoning_incorrect_3,inconsistency_llm_reasoning}. Furthemore, they are often accompanied by confidence-like certainty indicators (sometimes unjustified)~\citep{llm_uncertainty}.} As a result, such rationales can increase user trust and confidence without improving the quality of model outputs, potentially drawing users' attention away from independent fact verification in real-world decision-making scenarios under AI assistance.



\begin{figure*}[!h]
    \centering
    \includegraphics[width=0.980\linewidth]{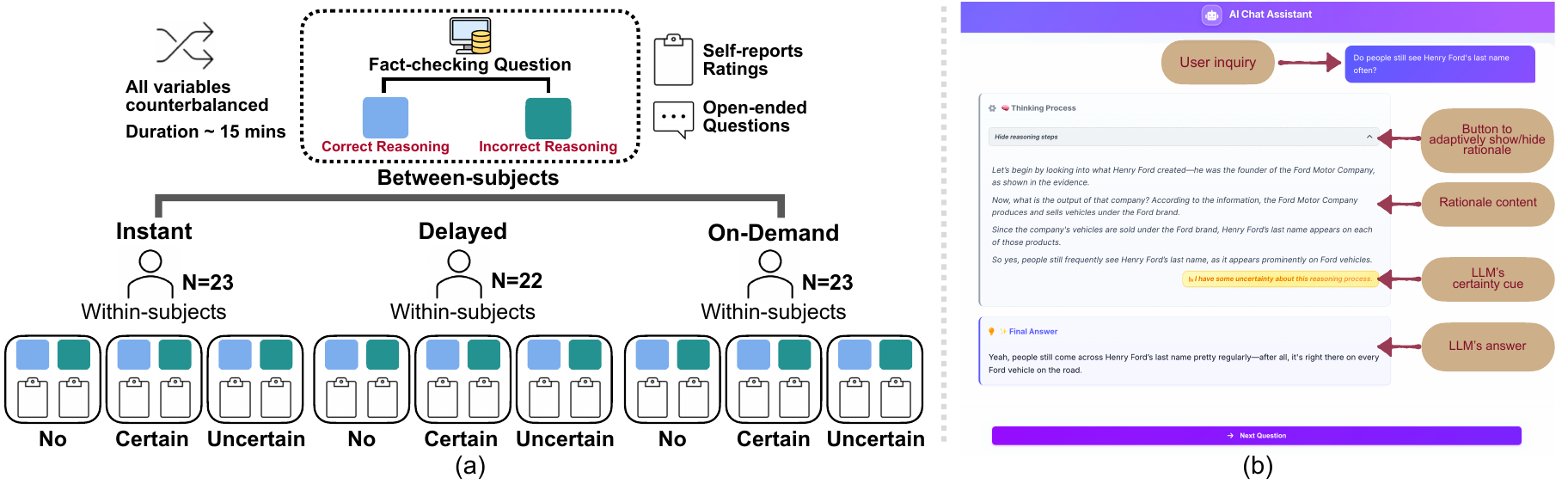}
    \vspace{-2.2mm}
    \caption{
    \revise{
    \emph{(a)} Study procedure and design: participants were assigned to one rationale \emph{presentation format} (Instant/Delayed/On-demand; between-subjects) and completed six trials covering all combinations of \emph{correctness} (correct/incorrect) and \emph{certainty cue} (none/certain/uncertain; within-subjects), with counterbalanced order. 
    \emph{(b)} Web interface: the query, LLM answer, and rationale were shown according to the assigned format, with the certainty cue when applicable.}
    }
    \vspace{-1.4mm}
    \label{fig:procedure_interface}
\end{figure*}


Given the foregoing, we investigate LLM \emph{user-facing} rationales, the step-by-step justifications shown with LLM's answer, and ask how such rationales should be presented for better AI-assisted decision support. 
\revise{We focus on three factors: (IV1) \textbf{presentation format} (when/how the rationale is revealed), (IV2) \textbf{correctness} (whether the rationale is consistent with the answer), and (IV3) \textbf{certainty framing} (how confident the rationale sounds).}
We thereby ask:
\begin{enumerate}[label=, leftmargin=-0pt, itemsep=0pt]
    \item \textit{\textbf{RQ:} 
    \revise{How does \emph{presentation format, correctness, and certainty framing} of the LLMs' reasoning rationale influence users' trust and decision-making behaviors (i.e., advice adoption and decision confidence)?}}
\end{enumerate}



To address this, we ran an online study ($N$=68) that manipulated presentation format (instant vs.\ delayed vs.\ on-demand), correctness (correct vs.\ incorrect), and certainty cue (none vs.\ certain vs.\ uncertain). 
We aim to examine how the presentation and how confident the rationale was framed can jointly affect users' trust perceptions, adoption of AI advice, and decision confidence.

Our key findings showed that correctness and certainty framing are high-impact levers: \emph{correct rationales} and \emph{certainty cues} increased trust, decision confidence, and advice adoption, whereas \emph{incorrect rationales} and \emph{uncertainty cues} reduced them. 
Notably, certainty framing shifted trust and decisions even with the same rationale content, indicating effects beyond correctness. 
In contrast, the presentation format showed no significant differences. 
These findings highlight a risk: confidence framing can amplify persuasion and miscalibrate trust when it is not aligned with actual reliability.
Open-ended responses suggest participants used rationales mainly to \emph{audit and calibrate trust}, despite trust dropping sharply when rationales were inconsistent or mismatched with the answer itself. 
They further emphasized a need for rationales that are \emph{step-by-step and auditable}, \emph{grounded in verifiable evidence}, and \emph{adaptive}, including explicit uncertainty and self-correction, with summary-first presentation and on-demand expansion.




Overall, our work shows that LLM rationales are double-edged: they can support decision-making, but can also miscalibrate trust when weak reasoning is paired with strong certainty. We offer design guidance: present stepwise rationales that are easy to check, tied to evidence, and adjustable on demand, and communicate certainty and self-corrections in ways aligned with reliability to support appropriate trust in LLM-assisted decision support.

\section{Related Work: Effects of AI Explanations on User Trust and Decision-Making}

Several prior works in explainable AI (XAI)~\cite{xai} show that AI's explanations can influence users’ trust and decision-making~\citep{confidence_effect_in_decision_1,confidence_effect_in_decision_2}, reliance~\citep{explanation_ai_overreliance}, and mental models~\citep{impact_xai_decision_making}, which lead to decision biases~\citep{impact_xai_decision_making} and increase the acceptance of incorrect AI advice~\citep{Chen_Jing}. 
These findings highlight a key concern: explanations can influence users’ trust and decisions without improving model quality.
Recent large reasoning models (LRMs)~\citep{lrm_survey} change what ``explanations'' look like. 
Classic XAI typically provides model-derived explanations for a fixed prediction (e.g., feature attributions such as LIME~\citep{xai_shape_lime} and SHAP~\citep{xai_shape} or visualizations~\citep{xai}), whereas LRMs often produce free-form, natural-language rationales that resemble step-by-step human-like thinking process. 
Such rationales can be persuasive yet unfaithful to the model’s actual decision process, including being inconsistent with the final answer or containing incorrect reasoning~\citep{reasoning_incorrect_1,reasoning_incorrect_2,reasoning_incorrect_3}. 
This creates risks for trust calibration because users may treat the plausible rationales as evidence of quality.

Despite extensive model-centric work~\citep{lrm_survey} on reasoning techniques such as chain-of-thought~\citep{cot} for improving model performance, comparatively less is known about the \emph{user-facing} effects of showing these LLMs' reasoning rationales on users' trust and decision-making, especially when these rationales can be incorrect or inconsistent but with strong expressions of model certainty~\citep{llm_uncertainty}. 
Our work addresses this gap by testing how the presentation and correctness with certain framing of the user-facing LLM reasoning rationale influence users' trust and decision-making. 



\begin{table*}[ht!]
\centering
\footnotesize
\renewcommand{\arraystretch}{1.30}
\caption{An example of the factual verification question and LLM response shown to participants: answer and rationales (correct or incorrect, depending on condition) and an accompanying certainty cue (none/certain/uncertain, depending on condition).}

\vspace{-1.0mm}

\begin{tabularx}{\textwidth}{>{\raggedright\arraybackslash}p{0.256\textwidth} >
{\raggedright\arraybackslash}X} 
\hline
\multicolumn{1}{l}{\textbf{Component}} & \multicolumn{1}{l}{\textbf{Content}} \\
\hline

\textbf{Factual Verification Question\newline(User Inquiry)} & 
Is Miami a city on the American West Coast?
\\ \hline

\textbf{LLM's Answer} & 
No, Miami is not a city on the American West Coast. 
It's located in Florida, which is part of the southeastern region of the United States and lies along the East Coast.
\\ \cline{1-2}

\textbf{LLM's Reasoning Rationale\newline(Correct)} &  
First, let’s identify where Miami is located. Miami is a city in the state of Florida, and it sits in the southeastern part of the United States along the Atlantic Ocean.\newline
Now, let’s consider what is meant by the American West Coast. This term refers to the coastal states bordering the Pacific Ocean—California, Oregon, Washington, Alaska, and sometimes Hawaii. Since Florida does not border the Pacific Ocean and is located on the East Coast, it is not part of the West Coast.\newline
Therefore, Miami is not a city on the American West Coast.
\\ \cline{1-2}

\textbf{LLM's Reasoning Rationale\newline(Incorrect)} &  
Miami is located in the state of California, which is one of the primary West Coast states. (Here is the factual incorrectness in this step)\newline
Despite that, when we review the actual location of Miami, it's evident that it’s not on the Pacific Ocean but on the Atlantic, which confirms it’s not a city on the West Coast.\newline
In the end, we can conclude that Miami is not on the American West Coast.
\\ \cline{1-2}

\textbf{Certainty Cue} &  
No cue: no framing about model's certainty at all\newline
Certain cue (an example): I feel very certain about how I think about this question.\newline
Uncertain cue (an example): I'm somewhat hesitant about how I think about this question.
\\ \hline
\end{tabularx}
\vspace{-1.2mm}
\label{stimuli_example}
\end{table*}


\section{Methods}

\subsection{Study Design and Procedure}

We ran an online experiment with a mixed design: 2 (Rationale Correctness: correct vs.\ incorrect; within-subjects) $\times$ 3 (Certainty Cue: none vs.\ uncertain vs.\ certain; within-subjects) $\times$ 3 (Rationale Presentation: instant vs.\ delayed vs.\ on-demand; between-subjects). 

\textbf{Rationale presentation.} In the \textit{instant} condition, the rationale appeared with the answer; in the \textit{delayed} condition, it appeared after a brief, length-proportional delay following~\cite{response_delay}; and in the \textit{on-demand} condition, it was hidden by default and revealed via a \emph{Show/Hide thinking steps} button (Fig.~\ref{fig:procedure_interface}).

\revise{
\textbf{Certainty cues.} We manipulated the model's certainty on their reasoning by adding a one-sentence cue at the end of the rationale. In the \textit{certain} condition, the cue conveyed high certainty (e.g., ``I am confident in my reasoning''); in the \textit{uncertain} condition, it conveyed low confidence (e.g., ``I am not fully sure about my thinking process''); and the \textit{none} condition included no cue. 
}

Each participant completed six trials covering all combinations of Rationale Correctness and Certainty Cue. The trial order was counterbalanced. 
Participants consented, completed a pre-survey, and were randomly assigned to one presentation-format condition. 
In each trial, they entered a given factual query, reviewed the LLM’s answer and reasoning rationale, and completed a questionnaire after each trial. 
A final post-survey collected two open-ended responses. 
The procedure is shown in Fig.~\ref{fig:procedure_interface}.


\subsection{Materials}


\subsubsection{Web Interfaces and Stimuli}

We developed a custom web interface (Fig.~\ref{fig:procedure_interface}) that simulates conversational information seeking. 
In each trial, participants entered a given factual query and then viewed the LLM’s binary answer with its reasoning rationale and certainty cue (when applicable) according to the condition. 

\revise{
We used factual verification claims from a public fact-checking dataset~\citep{strategyqa} with model-generated stepwise reasoning rationales. 
We selected claims to be comparable in format and difficulty (short claims requiring evidence-based verification) and to avoid niche topics that would strongly depend on specialized knowledge. 
We counterbalanced trial order so any residual claim-specific differences would not systematically favor a condition. 
}

For each claim, we prepared two rationales while keeping length and structure comparable: a \textit{correct} rationale aligned with the answer from dataset, and an \textit{incorrect} rationale containing a controlled logical or factual flaw to the answer. 
Examples are shown in Table~\ref{stimuli_example}.


\subsubsection{Self-reported Measures}
After each trial, participants completed the following measures.

\revise{
\textbf{Decision (advice adoption):} whether the participant will \emph{accept or reject the LLM’s answer} (i.e., advice adoption). 
}

\textbf{Decision confidence:} 7-point Likert scale~\cite{confidence_rating} (i.e., ``How confident are you with the decision made?'').

\textbf{Trust in information:} perceived trust in each LLM's response, 5-point Likert scale~\cite{trust_info_1,trust_info_2} (e.g., ``Is this information trustworthy?'').

\textbf{Trust in LLMs:} overall trust in LLM system using 5-point Likert scale~\cite{measure_trust_system}.
\revise{This item captures perceived trust of the \emph{system as a whole} across tasks (i.e., willingness to rely on the LLM in general), beyond the correctness of any single response.}

\textbf{Manipulation check:} whether participants perceived the rationale as correct and consistent with the answer.


\subsection{Participants and Data Analysis}

A priori G*Power~\cite{gpower} analysis suggested 48 participants to detect a medium effect ($f=0.25$) with 90\% power. 
We recruited 68 English-speaking participants via Prolific~\cite{Prolific}. 
The study was approved by institutional ethical committee at the University of Amsterdam. 
Participants were compensated according to platform standards. 
Demographic details are reported in Table~\ref{table:online_demographics}.



\begin{table}[!h]
\centering
\small
\renewcommand{\arraystretch}{0.860}
\caption{Characteristics of participants in the online study.}
\vspace{-2.0mm}
\begin{tabularx}{\columnwidth}{>{\raggedright\arraybackslash}p{0.30\columnwidth} >{\raggedright\arraybackslash}p{0.40\columnwidth} >
{\raggedright\arraybackslash}X} 

\toprule
\textbf{Demographic} & \textbf{Category} & \textbf{N (\%)} \\
\midrule
Gender  & Female  & 33 (48.5\%) \\
        & Male    & 35 (51.5\%) \\
\midrule
Age     & 18--24  & 2 (2.9\%)  \\
        & 25--34  & 28 (41.2\%) \\
        & 35--44  & 20 (29.4\%) \\
        & 45--54  & 11 (16.2\%) \\
        & 55--64  & 5 (7.4\%)  \\
        & 65+     & 2 (2.9\%)  \\
\midrule
Education & Bachelor's degree    & 45 (66.2\%) \\
          & Master's degree      & 22 (32.4\%) \\
          & Doctorate or higher  & 1 (1.5\%)  \\
\midrule
AI Familiarity & Never      & 1 (1.5\%)  \\
               & Slightly   & 7 (10.3\%) \\
               & Moderately & 19 (27.9\%) \\
               & Very       & 32 (47.1\%) \\
               & Extremely  & 9 (13.2\%) \\
\bottomrule
\end{tabularx}
\vspace{-1.6mm}
\label{table:online_demographics}
\end{table}

\begin{table*}[!ht]
\centering
\renewcommand{\arraystretch}{0.96}
\small
\setlength{\tabcolsep}{4pt}
\caption{\emph{Descriptive statistics (online study).} 
Means (SD) for each measure by rationale format, certainty cue, and correctness.
}
\vspace{-1.6mm}

\begin{tabularx}{\textwidth}{@{}>{\raggedright\arraybackslash}X
  *{3}{>{\centering\arraybackslash}p{1.6cm}}
  *{3}{>{\centering\arraybackslash}p{1.4cm}}
  *{2}{>{\centering\arraybackslash}p{1.7cm}}@{}}
\toprule
& \multicolumn{3}{c}{\cellcolor{gray!20}\textbf{Rationale Format}} 
& \multicolumn{3}{c}{\cellcolor{blue!10}\textbf{Certainty Cue}} 
& \multicolumn{2}{c}{\cellcolor{red!10}\textbf{Rationale Correctness}} \\
\cmidrule(lr){2-4}\cmidrule(lr){5-7}\cmidrule(lr){8-9}
\textbf{Dependent Variable} 
  & \textbf{Instant} & \textbf{Delayed} & \textbf{On-demand} 
  & \textbf{None} & \textbf{High} & \textbf{Low} 
  & \textbf{Correct} & \textbf{Incorrect} \\
\midrule
Trust (Information)
  & 3.92 (0.54) & 3.91 (0.82) & 4.03 (0.78)
  & 3.97 (1.01) & 4.14 (0.90) & 3.75 (1.14)
  & 4.09 (0.98) & 3.81 (1.07) \\
Trust (LLM system)
  & 3.70 (0.68) & 3.83 (0.98) & 3.74 (0.97)
  & 3.76 (0.99) & 3.95 (1.02) & 3.55 (1.13)
  & 3.87 (1.04) & 3.64 (1.07) \\
\midrule
Decision (advice adoption)
  & 0.89 (0.17) & 0.95 (0.11) & 0.88 (0.18)
  & 0.91 (0.29) & 0.96 (0.21) & 0.84 (0.37)
  & 0.94 (0.24) & 0.86 (0.35) \\
Decision Confidence
  & 5.66 (1.13) & 5.60 (1.00) & 5.59 (0.63)
  & 5.59 (1.31) & 5.96 (1.29) & 5.31 (1.59)
  & 5.77 (1.39) & 5.47 (1.46) \\
\bottomrule
\end{tabularx}
\vspace{0.4mm}
\label{tab:survey_descriptive}
\end{table*}


\begin{figure*}[!htp]
    \centering
    \includegraphics[width=0.992\linewidth]{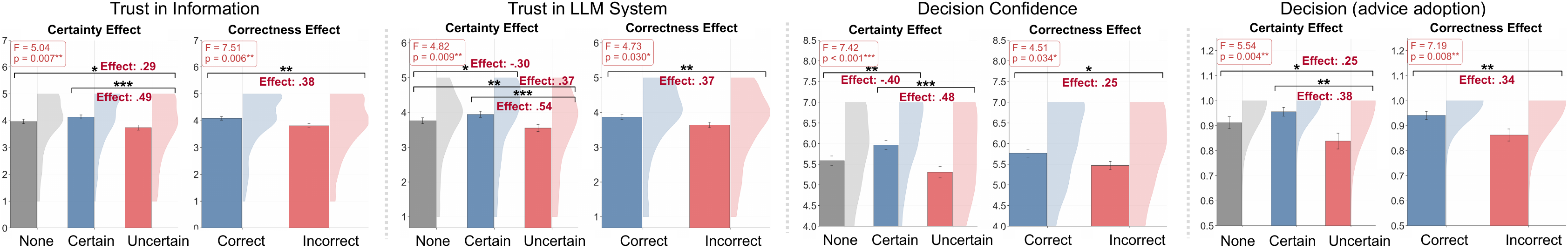}
    \vspace{-2.0mm}
    \caption{
    \revise{
    \emph{Main effects of certainty cues and rationale correctness.} 
    In each sub-figure, the left panel shows the certainty-cue effect and the right panel shows the correctness effect. 
    The y-axis shows the mean rating for the corresponding measure (Likert-scale score for trust and confidence; proportion for advice adoption decision).
    Bars are means; half-violins show distributions; The upper-left of each sub-figure $(F,p)$ are ANOVA results; brackets are post-hoc comparison tests (*$p<.05$, **$p<.01$, ***$p<.001$).
    }}
    \label{fig:survey_posthoc}
    \vspace{-0.2mm}
\end{figure*}


For data analysis, we conducted mixed ANOVAs on quantitative data to test the effects of three independent variables, followed by post-hoc pairwise comparisons, after checking the normality (Shapiro-Wilk~\cite{SHAPIRO1965}) and homogeneity of variance (Bartlett’s test~\cite{Arsham2011}). 
Open-ended qualitative responses were analyzed with inductive thematic analysis by two coders following~\cite{elo2008qualitative}.

\section{Quantitative Findings} 

Table~\ref{tab:survey_descriptive} reports descriptive statistics and Fig.~\ref{fig:survey_posthoc} shows statistic tests. 
Overall, trust and decision-making were influenced significantly by \textbf{rationale correctness} and \textbf{certainty framing}, while \textbf{presentation format} had little impact, addressing our RQ.

\textbf{Trust in information.} We found main effects of certainty cue ($p=.007$) and rationale correctness ($p=.009$). Post-hoc tests showed that \emph{uncertain} cues decreased trust relative to \emph{certain} ($p<.001$) and no cues ($p<.05$), and \emph{correct} rationales were trusted more than \emph{incorrect} ones ($p<.01$).


\textbf{Trust in LLM system.} There was a main effect of certainty cue ($<p=.009$). 
Certain cues increased trust in LLMs compared to uncertain cues ($p<.001$) and no cue ($p<.05$).
No cue was rated higher than uncertain cues ($p<.01$). LLM with \emph{correct} rationales were trusted more than \emph{incorrect} ones ($p<.01$).


\textbf{Decision confidence and advice adoption.} Certainty cues increased decision confidence and advice adoption: certain cues yielded higher confidence than uncertain cues and no cue. 
Correct rationale increased confidence and advice adoption more than incorrect ones ($<p=.05$;$<p=.01$).

\section{Qualitative Findings} 

We analyzed participants’ responses to two open questions: \emph{(Q1)} how seeing AI’s ``thinking process'' affected trust and decisions, and \emph{(Q2)} which reasoning features they found most helpful. 
\revise{We derived four themes (two themes per question) from the thematic analysis.}


\subsection{Effects of Seeing the AI’s Reasoning Rationales on Trust and Decision Making}


\paragraph{\textbf{Theme 1:} Participants used rationales to calibrate trust by auditing the model’s logic.}
Participants rarely treated rationales as determinative evidence. Instead, they read them to judge whether the answer was trustworthy and reliable. 
Many described actively checking the reasoning steps: ``I read it and examined the rationalization behind it,'' and ``Seeing the AI’s thinking process helped me see how much I could trust it.'' 
They looked for whether the model considered relevant factors (e.g., ``if they did or did not consider certain things that I was thinking about'') and whether it omitted or ``hid'' context (``It shows me if the answer it provided is hiding any context.''). 
This auditing role was most salient when participants felt uncertain or lacked domain knowledge: ``if it was a topic I had little knowledge on, it helped me feel more confident,'' and ``I was unfamiliar with some concepts and relied on AI to give me explanations.'' 
Overall, participants used rationales as a calibration tool, to decide whether to rely, cross-check, or dismiss the AI output.



\paragraph{\textbf{Theme 2:} Rationale was double-edged, with a penalty for inconsistency.}
While many participants said rationales increased trust when the logic was easy to follow (``It made me trust the answer more because I could follow the logic''), some distrusted it because the reasoning looked wrong or incomplete: ``It solidified that it was thinking about the question wrong and missing things,'' and ``It kept making logical errors which made me distrust.'' 
A recurring damaging pattern was an answer-rationale mismatch: participants noticed when the ``thinking process seemed inconsistent, even when the model seemed to reach the correct answer,'' which they experienced as confusing and destabilizing (``the final answer seemed more right, while reasoning said something else [...], it made me more wary''). 
In short, showing rationales amplified trust when coherent, but eroded trust when contradictions or mismatches became visible.


\subsection{Helpful Features and Desired Improvements of the LLMs Reasoning Rationales}


\paragraph{\textbf{Theme 3:} ``Helpful reasoning'' is stepwise and audible.}
Participants strongly valued reasoning that functioned as a logic map rather than a narrative stream, prioritizing structural clarity over linguistic fluency. They specifically requested step-by-step reasoning ''that lays out separate facts'' individually before integrating them into a conclusion. 
This structured decomposition transformed the reasoning process from a passive reading experience into an active audit trail, enabling users to pinpoint exactly where errors occurred: ``Showing its working because then I could see where the bad information was coming from,'' and ``Breaking the solution/answer into steps so I can check at what step the answer is going in the wrong direction.'' 
Consequently, the utility of reasoning lay in its ``auditability'', by exposing discrete logical chain, users could verify if model has actually ''understood the prompt constraints and ensure the final output was connected to the text provided'' without internal contradictions. 
As one participant noted, this structure serves as a trust calibration: ``If I see that there is something wrong in one of the steps, I won’t trust the answer.''


\paragraph{\textbf{Theme 4:} Participants valued explicit uncertainty signaling, self-correction, and controllable reasoning depth.}
Beyond structural clarity, participants value model's display of metacognitive awareness. 
Several valued cues that conveyed how certain AI was (e.g., ``a summary of how certain AI was in its argument''), and reported that uncertainty statements helped them calibrate trust: ``explicit uncertainty statements allowed me to see how much I could trust the AI.'' 
Participants also appreciated visible self-correction as a form of quality control (e.g., ``Mostly when AI corrects itself and corrects its answer''), suggesting that updating or revising an earlier response can increase perceived trustworthiness compared to confidently sticking with a flawed conclusion.
Moreover, participants also emphasized control over reasoning depth. Some preferred compressed formats (``bullet point it easier and clearer to consume''), while others wanted the option to expand into details when needed (``Show the thinking process in summary then the full version''). Several explicitly requested user control and follow-ups (``Maybe the option to further question its outcomes''). 
Overall, participants described helpful rationales that both signal uncertainty when appropriate and support adjustable, on-demand inspection.

\section{Discussion}


\subsection{Rationales as a Calibration Interface, Not Just Transparency}

Our participants did not treat rationales as determinative proof, but used them to \emph{calibrate trust} by auditing the model’s logic. 
In open-ended responses, participants described ``rechecking each step'' and using the rationale to see ``whether AI considered certain things'' or ``hiding any context''. 
This echoes prior work that explanations shape users’ mental models and decision bias~\cite{impact_xai_decision_making}, while our findings highlight more specifically: LLM's rationales function as a \emph{calibration} tool, not merely a persuasive narrative.

Our results also show that effect is sensitive to \emph{reasoning quality and certainty framing}. 
Quantitatively (Fig.~\ref{fig:survey_posthoc}), \textbf{certainty cues} reliably increased \emph{trust} and \emph{decision confidence} compared to uncertainty cues, while \textbf{uncertainty cues} lowered trust below even no-cue baseline. 
Certainty framing also shifted participants' behavior, where they were more likely to follow advice under certainty than uncertainty cues. 
Correctness effects were smaller but still visible in trust. 
Participants’ open-ended responses mirror this asymmetry: confident, well-structured rationales were described as ``convincing'' and ``more trustworthy'', whereas ``unsure responses'' and ``logical errors'' triggered doubt. 
These findings echo the prior work of models' confidence on trust~\cite{confidence_effect_in_decision_1,llm_uncertainty,measure_confidence}, highlighting that participants treated LLMs' rationales as a calibration of trust and a signal for verification, rather than determinative proof for decision-making.

\textbf{Practical implications.} 
Systems or the LLM-powered UIs should design rationales as \emph{verification scaffolds}, not persuasive text: 
(a) present reasoning in stepwise, auditable units so users can trace what was considered and pinpoint where it goes wrong as participants wanted to ``see where wrong information was coming from'';  
(b) tie reasoning steps to \emph{specific supporting evidence} (e.g., quoted spans or highlighted excerpts from supporting context), reflecting repeated requests for provenance; and
(c) use progressive presentation (cf., \cite{Springer2020}), a summary by default with on-demand expansion and enabling users' follow-ups to fit varied depth preferences without encouraging skimming-based overreliance.


\subsection{How Rationales Harm Trust: Inconsistent Reasoning and Overconfident Framing}

Our findings also show that rationales can actively harm trust when they reveal errors or inconsistencies. 
Quantitatively, incorrect rationales reduced trust significantly more than showing no rationale. 
Qualitatively, participants described distrust when the model’s reasoning rationale was ``wrong and missing things''. 
A particular failure mode was \emph{answer-rationale mismatch} defined by prior work as ``unfaithful'' reasoning~\cite{reasoning_incorrect_1,reasoning_incorrect_2,inconsistency_llm_reasoning}: participants reported the ``thinking process seemed inconsistent'' even when model ``reach correct answer''. This mismatch produced skepticism because it undermines the rationale’s role as a reliable audit function. 
These results extend prior work on explainable AI (XAI)~\cite{incorrect_xai_1,incorrect_xai_2,incorrect_xai_3,incorrect_xai_4}
shows that explanations can backfire: when an explanation is low-quality or contradicts the facts or advice the AI is meant to justify, users evaluate the system more negatively, reduce trust and acceptance, triggering a strong credibility penalty

Certainty framing further amplified such trust patterns. 
Our quantitative data show that certainty cues increased trust in both information and LLMs, and decision confidence relative to certainty cues, even when the rationale content was unchanged. Uncertainty framing decreased trust below the no-cue baseline. Participants suggest the reason that they read confident language and detailed reasoning as competence signals (e.g., ``trusted… when it was very detailed''), whereas ``unsure responses'' made them ``not trust AI''. 
Thus, certainty cues function as high-risk social/epistemic signals that can miscalibrate trust if they are not aligned with true reasoning quality, echoing the work~\cite{llm_uncertainty}.

\textbf{Practical implications.} 
Design rationales as verification supports rather than persuasive narratives.
Two safeguard recommendations follow from these findings. 
First, systems should prioritize \emph{consistency checks} before displaying rationales (ensuring the rationale does not contradict the final answer), since mismatch is disproportionately trust-damaging. 
Second, certainty framing should be treated as a controlled, calibrated signal: avoid strong certainty language for low-quality rationales, and prefer calibrated uncertainty communication (e.g., visually graded indicators) that informs without triggering blanket distrust. 
Together, these measures target the core failure modes our study uncovered: transparency can backfire from flawed reasoning and miscalibration that is driven by the model's certainty framing.

\vspace{-0.4mm}
\section{Limitations and Future Directions}
\vspace{-1.0mm}



Our study has several limitations that also point to concrete next steps in future work.

First, our setting was controlled factual verification with a fixed response per trial. While this made it possible to isolate the effects of rationale correctness, certainty framing, and presentation, real decision support is often open-ended and interactive. 
Future work should test whether the same trust shifts hold in multi-turn conversations and in tasks with multiple acceptable answers (e.g., planning, trade-offs), where users can probe, request sources, and challenge the model.

Second, we manipulated \revise{explicit} certainty framing as a surface cue, but users may also infer model confidence from \revise{implicit} language cues (e.g., verbosity and fluency). 
Future study should jointly vary explicit certainty cues and implicit signals such as hedging language and step granularity, and measure when these signals help calibration of trust and reliance.

Third, participants repeatedly asked for reasoning that is auditable and controllable (e.g., stepwise, summary-first with on-demand expansion), yet our study only provided a simple show/hide option and did not manipulate reasoning depth. 
\revise{Future work could evaluate more advanced on-demand designs that compare layered summaries against full rationales and evaluate step-level evidence linking as well as follow-up auditable interactions to see which best supports verification without increasing cognitive load.}


Finally, \revise{we caution that our ``rationales as a trust calibration interface'' framing should be understood as an actionable design lens grounded in observed user behaviors, rather than a new theory of trust.}
A concrete next step is to operationalize this lens in adaptive interfaces that adjust \emph{what} is shown based on behavioral signals, for example, surfacing sources or verification prompts when users' confidence is high but evidence attention is low, or offering more structured, stepwise reasoning when users appear uncertain (e.g., via interaction traces such as dwell time, scrolling, and clicks). 
Testing whether such adaptive interfaces improve calibrated trust and reliance over time is also an important next step.

\section*{Appendices}


\begin{appendices}

\setcounter{table}{0} 
\setcounter{figure}{0} 


\section{Ethical Statement}
\label{sec:appendix_ethics}

We carefully considered the ethical implications of this research, including the risks of handling sensitive user information and potential biases in generated content. 
This study adheres to the ethical standards of our institute, with strict measures to safeguard participant data and prevent harmful or biased AI outputs. 
Informed consent was obtained from all participants, ensuring anonymity and the option to withdraw at any time without repercussions. 


\section{AI Usage Disclosure}
\label{sec:appendix_ai_disclosure}

We used AI tools (i.e., GPT-5.2) only for language editing to improve clarity and conciseness.
All study design, analysis, literature review, and writing were conducted and verified by the authors.










\end{appendices}



\begin{acks}
This work was supported by the JST CREST Grant (JPMJCR2562), JST K Program Grant (JPMJKP24C2), and JST FOREST Grant (JPMJFR232R) in Japan.
\end{acks}

\bibliographystyle{ACM-Reference-Format}
\bibliography{chiea26-408-1,chiea26-408-2}


\end{document}